\documentclass[aip,graphicx]{revtex4-1}

\usepackage{graphicx}
\usepackage{dcolumn}
\usepackage{bm}

\usepackage[utf8]{inputenc}
\usepackage[T1]{fontenc}
\usepackage{mathptmx}
\usepackage{etoolbox}

\usepackage{graphicx}
\usepackage{natbib}
\usepackage{xfrac}
\usepackage{comment}
\usepackage{siunitx}
\usepackage[a-1b]{pdfx} 
\usepackage{amsmath}
\allowdisplaybreaks
\usepackage{subcaption}
\usepackage{caption}
\usepackage{xcolor}
\usepackage{siunitx}

\makeatletter
\def\@email#1#2{%
 \endgroup
 \patchcmd{\titleblock@produce}
  {\frontmatter@RRAPformat}
  {\frontmatter@RRAPformat{\produce@RRAP{*#1\href{mailto:#2}{#2}}}\frontmatter@RRAPformat}
  {}{}
}%
\makeatother

\draft 

\begin{document}


\title{Novel Approximation of the Modified Mild Slope Equation } 



\author{Cheng-Nian Xiao}

\email{chx33@pitt.edu}

\affiliation{Center for Research Computing and Data, University of Pittsburgh, Pittsburgh, PA 15261, USA.}



\begin{abstract}

The mild-slope equation and its various modifications aim to model, with varying degrees of success,  linear water wave propagation over sloping or undulating seabed topography.  However,  despite multiple modifications and attempted simplifications, the different variants of the equation include multiple higher order  terms involving the nonlinear water wave dispersion relation and thus remain analytic intractable. To further facilitate its use, we derive a drastically simplified alternative version of the modified mild-slope equation that bears striking resemblance to the linear shallow water equation while retaining all critical features of the original equation that enable it to be valid for a wide range of wave numbers and water depths.  Direct comparison of the modified mild-slope equation and our simplified formulation  indicates that the simplified equations  agree with the modified mild slope equation at leading orders in the forcing frequency and local depth. Validations with multiple sets of benchmark wave scattering problems demonstrate that despite the clearly reduced complexity, the simplified equations were able to replicate the predictions of the modified mild slope equations over a wide range of wave numbers and surface topographies, including  higher order resonant conditions.\\

Topics: Mild Slope Equations, Shallow Water Equation, Surface Wave Scattering, Undulating Topography, Bragg resonant reflection 

\end{abstract}


\maketitle 

\section{Introduction}

The study of surface wave scattering by seabed topography is vitally important for  predicting coastal processes to help marine structures design, wave power extraction and habitat preservation, while also providing fundamental insights into wave energy dissipation and scattering mechanisms\cite{dean1991water,dingemans1997water,li2007bragg,garnaud2010bragg}.
The dynamics of  water waves can be adequately described by Laplace’s equation in combination with other equations specifying boundary and/or radiation conditions\cite{mei2005theory}. However,  unless the seafloor is flat, it is extremely challenging to find exact analytical solutions for this type of problems, and they typically only apply to simple geometries with horizontal or vertical boundaries.
To overcome this limitation, approximations  such as the linear shallow-water equations(SWE) or mild-slope equations(MSE) have been developed. These approaches simplify the problem by approximating the vertical dependence of the flow through depth-averaging of the original linear wave equations, thus making it feasible to study wave scattering over slowly varying topography\cite{friedrichs1948waves,stoker2019water}. 
However, the SWE is only an accurate model when restricted to very long waves with respect to depth, such as tides or tsunamis, due to its neglect of depth-dependent wave dispersion relations. Simulations of realistic coastal wave dynamics instead rely on Boussinesq-type models which systematically incorporate higher order dispersive effects\cite{peregrine1967long,brocchini2013reasoned,chen2006fully,shi2012high,liu2019numerical,xiong2025improved}, but at the cost of significantly greater mathematical and numerical  complexity compared to the simplified SWE framework. 
On the other hand, the MSE, as originally discovered independently by \citet{berkhoff1973computation,berkhoff1976mathematical} and \citet{smith1975scattering} among others, aimed to overcome the limitations of the SWE and has undergone multiple modifications and improvements since its inception \cite{kirby1986general,miles1991variational,chamberlain1995modified,porter2003mild,porter2020mild,hsu2003bragg}.
Of particular note are the so-called complementary mild-slope equations(CMSE) derived by \citet{kim2004new}, the extended mild-slope equations(EMSE) by \citet{kirby1986general},  and modified slope equation(MMSE) introduced by \citet{chamberlain1995modified} which improve the original formulation of MSE \cite{berkhoff1973computation} for short wavelengths and in the presence of moderate to strong surface gradients such as sinusoidal  ripples \cite{davies1984surface} or the resonant Bragg reflection by artificial bars\cite{kirby1990bragg}. The MMSE  has found widespread success in application to surface wave scattering over sloping and undulating topographies\cite{chamberlain1995modified,chamberlain1995decomposition,liu2012analytic,liu2013analytic,liu2020bragg, liu2025zero}.   Due to its analytic complexity, the MMSE itself has also undergone multiple modifications and simplifications \cite{porter2003mild,liu2014explicit} to facilitate application.
In particular, a  simplified version of the MMSE has been proposed by \citet{porter2019extended} that amounts to an extension of the linearized shallow water equation to apply to variable depth topography and short waves. However, none of the above variants of the MMSE were able to attain both analytic simplicity with accuracy over a wide range of conditions. For example, the extended linear shallow water equations by \citet{porter2019extended} were shown to be less accurate at large surface slope values, whereas even the simplified versions of MMSE\cite{porter2003mild,porter2020mild} still require the solution of the nonlinear wave dispersion relation as well as the  higher order derivatives of the surface topography at each point.\\

This work introduces a remarkable simplification of the MMSE \cite{chamberlain1995modified,porter2003mild} which further reduces its computational and numerical complexity by replacing the nonlinear implicit local wave dispersion relation with an explicit approximate expression that does not contain higher derivative terms of the bedform $h(x,y)$. We demonstrate how the new equations can be elegantly reformulated to retain the familiar structure of linear SWE. This transformation offers two key advantages: Firstly, it retains the accuracy of the original MMSE over a wide range of wave-lengths, and  secondly, it preserves the  simplicity of standard SWE - facilitating theoretical analysis and numerical implementation  based on existing SWE frameworks with negligible additional cost.

\section{The Simplified Modified Mild Slope Equation}

\subsection{The Modified Mild Slope Equation }

Our setup assumes  incompressible, homogeneous and irrotational fluid motion over a bed of varying quiescent depth \( h(x,y) \), where  \( x \) and \( y \) denote the horizontal Cartesian coordinates.  \( z \)  is the vertical coordinate with the undisturbed free surface at \( z=0 \) and is positive for upward locations.

If the system is subject to forcing given by the velocity potential \( \Phi \) with angular frequency  $\omega$ as below: 
\[
\Phi(x,y,z,t) = \text{Re}\left( \phi(x,y,z) e^{-i\omega t} \right),
\]
 then according to linearized wave theory,  \( \phi \) must satisfy\cite{mei2005theory}  
\begin{align}
\nabla^2 \phi &= 0 \quad (-h < z < 0),  \label{eqnphi1}\\
\phi_z - \frac{\omega^2}{g} \phi &= 0 \quad (z=0),  \label{eqnphi2} \\
\phi_z + \nabla_h h \cdot \nabla_h \phi &= 0 \quad (z=-h), \label{eqnphi3}
\end{align}
where \( \nabla = (\partial/\partial x, \partial/\partial y, \partial/\partial z) \) and \( \nabla_h = (\partial/\partial x, \partial/\partial y) \) denote the total and horizontal gradient operators, respectively.
The free surface elevation is given by \( \xi(x,y,t) = \text{Re}\left( \eta(x,y) e^{-i\omega t} \right) \) where
\begin{equation}\label{eqnphi4}
\eta(x,y) = \frac{i\omega}{g} (\phi)_{z=0}.
\end{equation}

 \( \phi \) is subject to either lateral boundary conditions or a radiation condition in the case  of infinite extent, which will be of no further consideration for later part of this work.\\

The modified mild slope equations(MMSE) can be derived as a simplification and approximation for Eqn.  (\ref{eqnphi1})-(\ref{eqnphi4}) by  eliminating the dependence on the vertical coordinate $z$ through depth averaging and the application of suitable variational principles \cite{chamberlain1995modified,porter2003mild,miles1991variational}. They are formulated  in terms of the free surface elevation as follows \cite{chamberlain1995modified}:
 
 \begin{equation}\label{eqnmmse}
\nabla_h \cdot u_0 \nabla_h \eta + (k^2 u_0 + r)\eta = 0
\end{equation}

where the local wave number $k$ satisfies the dispersion relation 
\begin{equation}\label{eqnkh}
\frac{\omega^2}{g}=k\tanh(kh)
\end{equation}
dictated by the height $h=h(x,y)$ at each point. The term \( r(h)  \) has the explicit expression
\begin{equation}
r(h) = u_1(h) \nabla_h^2 h + u_2(h)(\nabla_h h)^2,
\end{equation}

with
\begin{align}
u_0(h) =&  \frac{1}{2k} \tanh(kh) \left( 1 + \frac{2kh}{\sinh(2kh)} \right) ,\\
u_1(h) =& \frac{\text{sech}^2(kh)}{4(K + \sinh(K))} \left\{ \sinh(K) - K \cosh(K) \right\},\\
u_2(h) =& \frac{k \, \text{sech}^2(kh)}{12(K + \sinh(K))^3}
\{ K^4 + 4K^3 \sinh(K) - 9 \sinh(K) \sinh(2K) \notag \\
&+ 3K(K+2\sinh(K))(\cosh^2(K) - 2\cosh(K) + 3) \},
\end{align}

where we have used  the abbreviation \( K = 2kh \). It has been noted that this formulation of MMSE can be transformed into the original mild slope equations by setting the term $r$ to zero at the expense of modeling accuracy  for undulating topographies\cite{chamberlain1995modified}.\\

\citet{porter2003mild} has demonstrated that the above formulation (\ref{eqnmmse}) for MMSE can be transformed into an equivalent equation of simpler structure as below:

\begin{equation}\label{eqnmmset}
\nabla \cdot k^{-2} \nabla \zeta + \{1 - v (\nabla h)^2\} \zeta = 0, 
\end{equation}

where  \(v\) is given by

\begin{equation}
v(h) = \frac{3(2K + \sinh K)(\sinh(2K) - \sinh K) - 3K^2(\cosh(2K) + 2) - 4K^3 \sinh K - K^4}{3(K + \sinh K)^4},
\end{equation}

and the free-surface elevation $\eta$ is related to \(\zeta\) of Eqn. (\ref{eqnmmset}) as 

\begin{equation}\label{eqnetazeta}
\eta(x, y) = \frac{2 \cosh(kh) \cdot \zeta(x, y)}{\sqrt{k(2kh + \sinh(2kh)}}.
\end{equation}
where it can be readily verified that in the shallow water limit $h \ll 1$, we have $\eta \approx \zeta$.

\citet{porter2003mild,porter2020mild} has shown that the term $v(h)$ is bounded as $0<v(h)<0.03$ and is far  smaller than its theoretical maximum value in most applications. Hence, the equivalent formulation of the MMSE given in Eqn. (\ref{eqnmmset}) can be significantly simplified by dropping the $v(h)$ term, which then leads to:

\begin{equation}\label{eqnmmses}
\nabla \cdot k^{-2} \nabla \zeta + \zeta = 0, 
\end{equation}
This equation can be readily seen as a generalized version of the  linear shallow water equations which are valid under the condition $kh\gg 1$ and for which the dispersion relation is $\omega^2/g=k^2h$. It incorporates the depth-dependent wave dispersion and is  shown to agree with the original MMSE (\ref{eqnmmse}) to the leading order, making a valid approximation beyond the long wave limit for many practical problems\cite{porter2003mild, porter2020mild}. Hence, we will use the term MMSE to refer interchangeably to either Eqn. (\ref{eqnmmse}) or (\ref{eqnmmses}) in  later parts of this work. However, Eqn. (\ref{eqnmmses}) still requires the evaluation of the wave number $k$ as a function of the local topography height $h(x,y)$ at every point $(x,y)$, which involves solving the nonlinear dispersion relation (\ref{eqnkh}) for water waves. In the following, we show how this equation can be further simplified by replacing $k(h)$ with an explicit expression that can be more readily computed and analyzed.

\subsection{Simplified formulation }\label{sec_sim2d}

Starting from the modified mild slope equation (\ref{eqnmmses}) introduced by \citet{porter2003mild}, our goal is to replace the  the local wave number $k$ in Eqn (\ref{eqnmmses}) by a simpler explicit expression. We do this by following two approaches: First, we  obtain the Taylor expansion  for $k(h)$ in terms of $h$, and then we derive a second expression which agrees with the Taylor series to leading order but leading to a simpler expression when substituted into (\ref{eqnmmses}). We further show that both of these resulting formulations can be regarded as generalizations of the linear shallow water equation, similar to those introduced in \citet{porter2019extended} but with simpler analytic structure.

Starting from the nonlinear dispersion relation $k=k(h,\omega^2/g)$ for water waves given by Eqn. (\ref{eqnkh}), we apply Langrange's theorem on Taylor expansion for inverse functions\cite{lagrange1770nouvelle} to obtain the following relation between the wave number $k$, water height $h$ and $\nu=\omega^2/g$:

\begin{equation}\label{eqnkh1}
    k(h,\nu)\approx k_1(h,\nu)=\nu^{1/2} \sqrt{ h^{-1} + \frac{1}{3}\nu + \dots }
\end{equation}

It can be shown through simple calculations that the following alternative  expression agrees with (\ref{eqnkh1}) to the leading order:

\begin{equation}\label{eqnkh2}
    k(h,\nu)\approx k_2(h,\nu)= \nu^{1/2}\left( h - \frac{1}{3}\nu h^2 + \dots \right)^{-1/2}
\end{equation}

Keeping the leading two orders of  (\ref{eqnkh1}) or (\ref{eqnkh2}) and substituting them into Eqn. (\ref{eqnmmses}) to replace $k$ hence results in two possible simplified versions of the modified mild slope equation  as given below:

\begin{equation}\label{eqnsmse1}
\nabla \cdot \left(\frac{\nu^{-1}h}{1 +\frac{1}{3} h\nu}  \nabla \zeta \right)+ \zeta = 0, \tag{SMSE1}
\end{equation}

\begin{equation}\label{eqnsmse2}
\nabla \cdot \left( (\nu^{-1}h-\frac{1}{3} h^2)  \nabla \zeta \right) + \zeta = 0, \tag{SMSE2}
\end{equation}

It can be readily seen that these two equations are generalizations of the linear shallow water  wave equation, which  is governed by the simple dispersion relation $\nu=k^2h$ at the long wave or shallow depth limit and can hence be written as 

\begin{equation}\label{eqnswe}
\nabla \left( \nu^{-1}h\cdot \nabla \zeta \right) + \zeta = 0, \tag{SWE}
\end{equation}

We can see that the second version given by (SMSE2) contains a simpler mathematical expression than the first version (SMSE1) derived directly from the Taylor expansion of the dispersion relation (\ref{eqnkh}). However, as can be seen from Eqn. (\ref{eqnkh2}), its range of validity is more restrictive since it requires that the forcing frequencies satisfy $\nu < 3/h $.  

\begin{figure}[t]
 \begin{subfigure}{0.49\textwidth}
	\centering
	\includegraphics[width=0.95\textwidth]{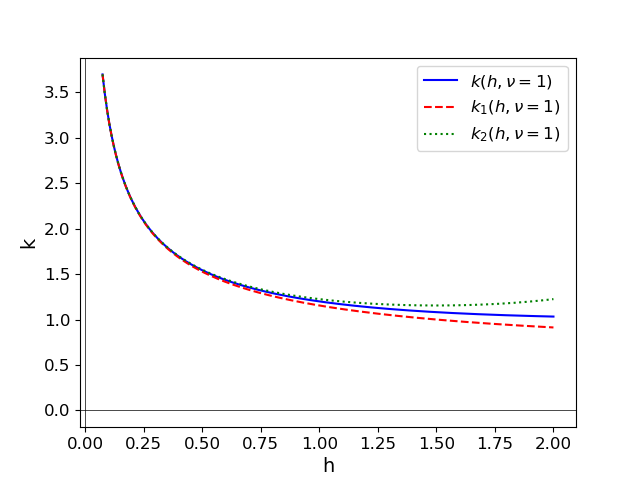}
   \caption{ }
	\label{subfig:kh_approx1}
 \end{subfigure}
  \begin{subfigure}{0.49\textwidth}
	\centering
	\includegraphics[width=0.95\textwidth]{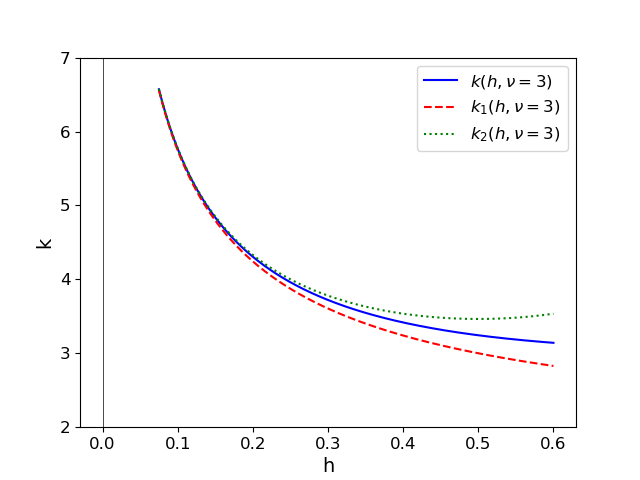}
   \caption{ }
	\label{subfig:kh_approx2}
 \end{subfigure} \\

	\caption{ Comparison of the two different simplified approximations $k_1,k_2$ given by (\ref{eqnkh1}) and (\ref{eqnkh2}) for the nonlinear wave dispersion relation  $k=k(h,\nu)$ satisfying $\nu=k\tanh(kh)$ for different temporal forcing frequencies $\nu=\omega^2/g$: (a) $\nu=1$, and (b) $\nu=3$.   }\label{fig:kh_approx}
\end{figure}

\begin{figure}

 \vspace{15pt}
   
		\centering
	\includegraphics[width=0.7\textwidth]{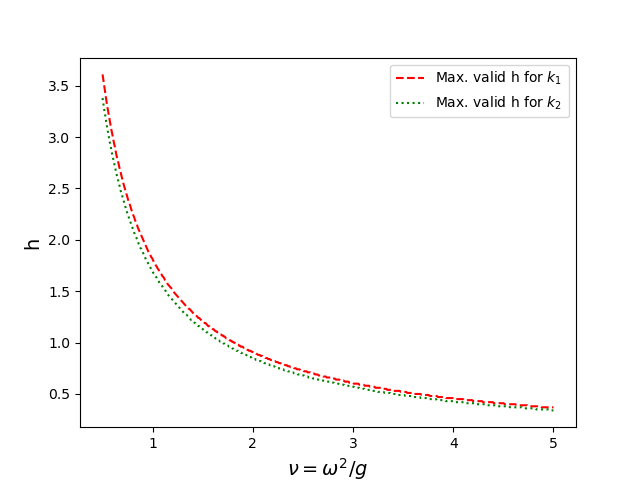}
 
	\caption{ Maximal height $h_{\text{max}}$ as a function of $\nu$ such that  for all $h< h_{\text{max}}$, the simplified  expressions (\ref{eqnkh1}) and (\ref{eqnkh2}) produce approximate  wave numbers $k_1,k_2$ that deviates less than 10\% from the exact value $k(h,\nu)$ given by $\nu=k\tanh(kh)$. }	\label{fig:fig_validh}
\end{figure}

In the following, we study the accuracy of each of these approximations compared to the exact dispersion relation (\ref{eqnkh}), which gives an indication over  which combination of wavelength and water depth the simplified equations will be accurate approximations to the slope equations given by Eqn. (\ref{eqnmmses}). In Fig. \ref{fig:kh_approx}, we compare both approximations  (\ref{eqnkh1}) and (\ref{eqnkh2}) with the exact wave number $k(h,\nu)$ determined by the dispersion relation $\nu=k\tanh(kh)$ for the two temporal forcing frequencies given by $\nu=\omega^2/g=1,3$. It can be seen by comparing Figs. \ref{subfig:kh_approx1} and \ref{subfig:kh_approx2} that  at the lower frequency $\nu=1$  both approximations are accurate over a larger range of water depth values  up until $h=1.75$, whereas for the higher frequency $\nu=3$, the range of validity for both schemes shrinks to water depths of  $0<h<0.35$. In Fig. \ref{fig:fig_validh}, we display the maximal water depth $h_{\text{max}}$ for which both approximations will be accurate to within 10\% of the exact value $k(h,\nu)$ as a function of the forcing frequency $\nu$. This confirms and quantifies the earlier observation that the range of validity for both simplified formulations decreases with growing forcing frequency $\nu$.  
This is not surprising, since, as a variant of the modified mild slope equation and a generalization of the shallow water equation, our formulation will be most accurate in the long-wave low-depth regime, similar as the other formulations of mild slope equations \cite{porter2019extended, chamberlain1995modified}.  We will show that this limitation does not prevent our proposed simplified equations from accurately modeling practical wave scattering problems that are most commonly studied in literature\cite{booij1983note,davies1984surface,kirby1990bragg}.  

\subsection{Determination of Wave Reflection and Transmission} \label{sec_r}

In the following, we only consider two-dimensonal water waves without variation along the horizontal $y$ direction. We assume that the water depth $h(x)$ is constant except within the limited region $0<x<L$:
\[
h(x) = 
\begin{cases} 
h_0 & \forall x \leq 0, \\
h_1 & \forall x \geq L,
\end{cases}
\]
where \( h_0 \) and \( h_1 \) are constant heights of the surface bed on each side of the varying topography, and \( L \) is the total length of the varying topography. Given the water depth function $h(x)$, we can solve either one of  the  mild slope equations (\ref{eqnmmse}), (\ref{eqnmmses}), (\ref{eqnsmse1}), (\ref{eqnsmse2}) for the free surface elevation $\eta(x)$ over $x \in [0,L]$. We then have to match the solution at the boundary points $x=0,L$ subject to continuity with the incoming and outgoing wave forms which are of the form:

\[
\eta(x) = 
\begin{cases} 
e^{ik_0x} + Re^{-ik_0x} & (x < 0), \\
Te^{ik_1x} & (x > L).
\end{cases}
\]

where $R,T$ are the reflection and transmission coefficients, respectively, and $k_0$ is the incident wave number directly related to the forcing frequency through $\nu=\omega^2/g=k_0\tanh(k_0\cdot h_0)$. A more detailed description of the implementation of this procedure to determine the reflection and transmission coefficients can be found in the work by \citet{chamberlain1995decomposition}.

\section{Validation}
\subsection{Roseau's Problem}
As one of the rare examples in linear water wave theory that possesses an exact analytic solution, we present Roseau's scattering problem \cite{roseau2012asymptotic} over a non-constant topography which is parameterized as $z(\xi)=-h(x(\xi))$ with $\xi \in (-\infty,\infty)$:

\begin{align}
x(\xi)/h_{0} =& \xi-(2\pi\beta)^{-1}(1-h_{L}/h_{0})\ln(1+\mathrm{e}^{2\beta\pi\xi}+2\mathrm{e}^{\beta\pi\xi}\cos(\beta\pi)) \label{eqnros1} \\
z(\xi)/h_{0} = & -1+(\pi\beta)^{-1}(1-h_{L}/h_{0})\tan^{-1}\{\sin(\beta\pi)/(\mathrm{e}^{-\beta\pi\xi}+\cos(\beta\pi))\} \label{eqnros2}
\end{align}

Thus $h^{\prime}(x(\xi))=-z^{\prime}(\xi)/x^{\prime}(\xi)$, and $\beta\in(0,1)$ is a parameter that determines the degree of shoaling. The ratio $h_L/h_0$ fixes the steepness of the topography. The wave reflection coefficient magnitude is then given by\cite{roseau2012asymptotic,porter2019extended}: 
\begin{equation}\label{eqnrR}
|R|=\left|\frac{\sinh[(k_{0}h_{0}-k_{L}h_{L})/\beta]}{\sinh[(k_{0}h_{0}+k_{L}h_{L})/\beta]}\right| 
\end{equation}

\begin{figure}

 \vspace{15pt}
   
		\centering
	\includegraphics[width=0.7\textwidth]{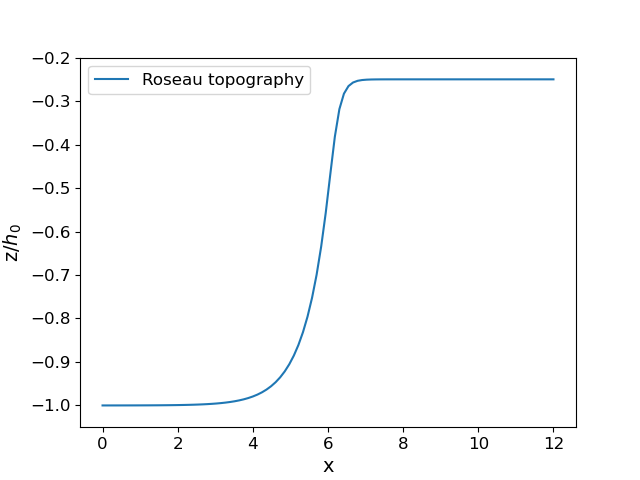}
 
	\caption{Limited portion of surface bedform for Roseau's problem parametrized by Eqns. (\ref{eqnros1}) and (\ref{eqnros2}). The geometric parameters are $\beta=0.5, h_L/h_0=0.25$. The $x$ coordinate has been shifted by 6 units to the right.}  	\label{fig:fig_roseaubed}
\end{figure}

This describes a topography that stays at nearly uniform depth for all $|x|\gg 1$ and undergoes nonlinear steepening in the middle section as determined by the parameters $h_L/h_0, \beta$. 
A limited portion of the  profile $z(x)$ for a Roseau bed topography with $\beta=0.5, h_L/h_0=0.25$ is shown in Fig. \ref{fig:fig_roseaubed} with a horizontal shift such that the left-most boundary of the bed is located at $x=0$.

\begin{figure}

 \vspace{15pt}
   
		\centering
	\includegraphics[width=0.7\textwidth]{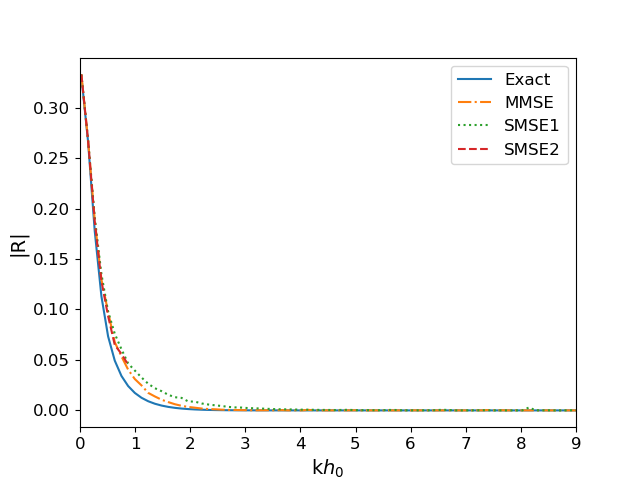}
 
	\caption{Reflection coefficient for the scattering problem over Roseau's profile as a function of the incident wave number $k$ normalized by the maximal height $h_0$. Comparison of numerical results obtained from  the simplified  modified mild slope equations with Eqn. (\ref{eqnmmses}) and the exact analytical value.  }	\label{fig:fig_roseau}
\end{figure}

To assess the fidelity of our simplified formulations   for the modified mild slope equations, we computed the reflection coefficients $R$ for different incident wave numbers $k$ based on the procedure in Sec. \ref{sec_r} using (SMSE1) and (SMSE2) and compared the results against both the exact value given by Eqn. (\ref{eqnrR}) as well as the numerical results obtained from the modified mild slope equation(MMSE). The formulation given by (SMSE2) is only valid for forcing frequencies $\nu<3/h$ and the corresponding incident wavenumbers $k$, so numerical results for (SMSE2) had to capped to $k$ values satisfying this constraint.   Our main purpose here  is to determine how accurately the simplified equations (SMSE1) and (SMSE2) approximate the MMSE given by Eqn. (\ref{eqnmmses}), which was shown to be nearly equivalent to the original MMSE in Eqn. (\ref{eqnmmse})\cite{porter2003mild}. As can be seen in Fig. \ref{fig:fig_roseau}, there is virtually no discernible difference between all three formulations except for the wave number range $1<kh_0<3$. It is especially encouraging to see that at the high frequency or steep slope limit, our simplified equation (SMSE1) was able to predict  the convergence of the reflection coefficient to zero, as dictated by the exact result (\ref{eqnrR}). This gives us confidence that our simplified formulations (SMSE1) and (SMSE2) are indeed an accurate representation of the MMSE over a wide range of wave numbers as encountered in practical problems.

\subsection{Booij's Problem}

\begin{figure}

 \vspace{15pt}
   
		\centering
	\includegraphics[width=0.7\textwidth]{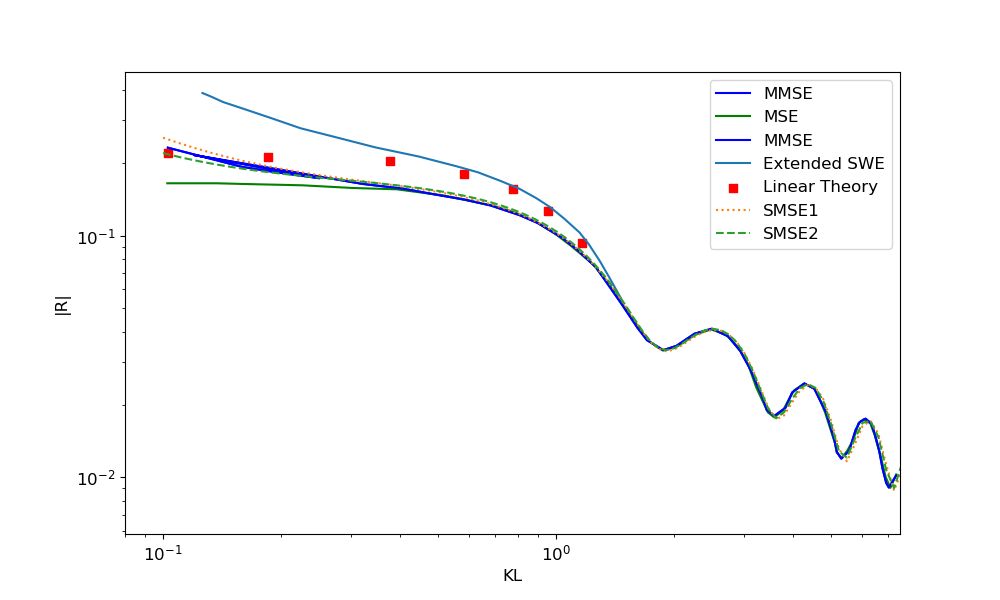}
 
	\caption{ Reflection coefficient for the scattering problem over planar slope (Booij) as a function of the incident wave number $k$ normalized by the length of steepening terrain $L$. Higher values of $L$ correspond to smaller slopes. Comparison of numerical results obtained from  the simplified  modified mild slope equations with other variants and the values from full linearized wave equations without depth-averaging. } 	\label{fig:fig_booij}
\end{figure}

A popular test problem that is widely used to assess the range of validity of different versions of  mild-slope equations as a function of slope magnitude is introduced in the work by \citet{booij1983note}.  In this configuration,  plane waves are scattered incident normally on a depth profile which is reduced to one third of its original height  through a planar slope of horizontal length $L$, thus \ $h_1 = h_0/3$ and $h(x) = h_0(1-2x/3L)$ for $x \in [0,L]$).  Booij  calculated the reflection coefficients with the help of  the original mild-slope equation developed by \citet{berkhoff1973computation} as well as from  the
linear wave equations (\ref{eqnphi1})-(\ref{eqnphi4}).
In the following, we compare Booij's data with numerical results of the reflection coefficient obtained through the following variants  of mild slope equations, which are: 1. Berkhoff's Mild Slope Equation(MSE)\cite{berkhoff1973computation}, 2. Porter's Modified Mild Slope Equation(MMSE) given by (\ref{eqnmmses})\cite{porter2003mild}, 3. Porter's extended shallow water equations (ESWE)\cite{porter2019extended}, 4. Simplified Mild Slope Equation based on approximation (\ref{eqnkh1}) for local wave number (SMSE1) and 5.  Simplified Mild Slope Equation based on  (\ref{eqnkh2}) for $k(h,\nu)$ (SMSE2).  
Fig. \ref{fig:fig_booij}
displays the  magnitude of the reflection coefficient $|R|$ as a function of  $k L$, where $k=k(h_0,\nu)$ is the incident wave number.  Booij’s results from solving the full linearized problem
are also shown as "ground truth" to validate the other numerical results. In the steep slope range with $kL<1$,  it can be seen that both simplified versions (SMSE1), (SMSE2) produce reflection coefficients that are nearly identical to those from the MMSE and are closer to  data points from the linearized equations, whereas both the original MSE and Porter's ESWE clearly show stronger deviations. This demonstrates that SMSE1 and SMSE2 are capable of handling steep slope cases just as MMSE and are a better generalization of the shallow water equations than MSE and ESWE while attaining a simpler mathematical structure.  

\subsection{Scattering over Sinusoidal Ripple Beds}

\begin{figure}

 \vspace{15pt}
   
		\centering
	\includegraphics[width=0.7\textwidth]{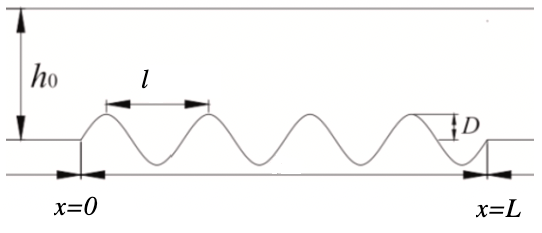}
 
	\caption{Sinusoidal ripple bed for wave scattering  experiment by \citet{davies1984surface}. Sketch reproduced from \citet{liu2019numerical} }	\label{fig:dh_bed}
\end{figure}

The ripple bed topography is another popular test case that has been extensively used to study the accuracy of different types of mild slope equations. It was well-known that both the linear shallow-water equation and wave scattering theory by \citet{miles1981oblique} as well as the original mild slope equation  were unable to predict the correct reflection coefficients  at the Bragg resonant conditions\cite{kirby1990bragg} for this type of topography,   which was the primary motivation behind the derivation of  more enhanced versions of mild slope equations \cite{chamberlain1995modified}.  In the following, we will evaluate the
accuracy of our simplified formulations (SMSE1), (SMSE2) for the modified mild-slope equation(MMSE) compared to the original MMSE version \cite{chamberlain1995modified,porter2003mild} at predicting scattering over such ripple bed topography. 
Experiments on water wave scattering have been carried out on a smoothly undulating bedform by  \citet{davies1984surface}   where the surface variation around a constant mean depth $h_0$ is given as $h(x)=h_0+\delta(x)$ with

\begin{equation}\label{eqnripple}
\delta(x) = D\sin(\ell x) \quad (0 \leq x \leq L),
\end{equation}

where \( h_0 \) is a constant and \( L = 2n\pi / \ell \). This means that the ripple topography is made of  \( n \) sinusoidal undulations of a single spatial frequency about the mean depth \( h_0 \). A sketch of this topography is displayed in Fig. \ref{fig:dh_bed}. Measurements for the reflection coefficient $R$ obtained in this experiment will be used to validate the accuracy of the simplified SMSE1, SMSE2 compared to the original MMSE.

\begin{figure}

 \vspace{15pt}
   
		\centering
	\includegraphics[width=0.7\textwidth]{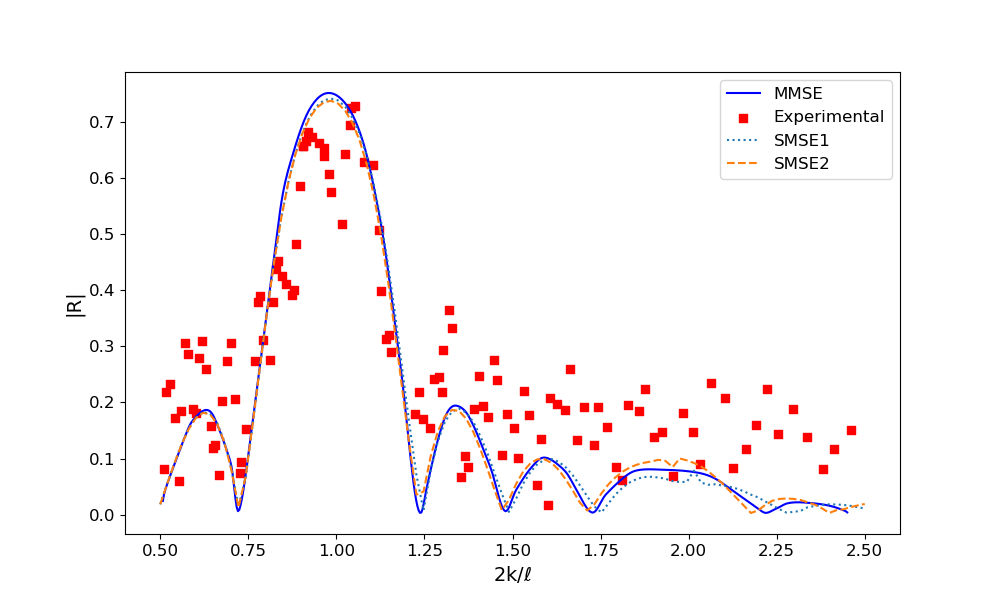}
 
	\caption{ Reflection coefficient magnitude $|R|$ of scattering over sinusoidal ripple bed as a function of the incoming wave number $k$ normalized by the wave number $\ell$ of the surface undulation. Experimental data by \citet{davies1984surface} for the case $D/h_0=0.32, n=4$ are compared to the MMSE and our simplified formulations (SMSE1), (SMSE2).} 	\label{fig:cpdh_comp}
\end{figure}

In Fig. \ref{fig:cpdh_comp}, the magnitude of reflection coefficient $|R|$ are plotted against $2k/\ell$, which represents
twice the ratio of the incident wavenumber  $k$ and the wave number $\ell$ of the bed topography as given in Eqn. (\ref{eqnripple}). For this particular case, the amplitude of the ripples relative to the mean depth is $D/h_0=0.32$, and the number of ripples in the bedform is $n=4$.   We compare the predictions obtained from solving (SMSE1) and (SMSE2) with those of MMSE and the experimental data. It can be seen that there is nearly perfect agreement between our simplified formulations and the original MMSE. In particular, besides correctly predicting the magnitude of the primary Bragg reflection at $2k/\ell=1 $, our numerical results from (SMSE1) and (SMSE2) were also able to capture the existence of higher order resonances at $2k/\ell \approx 2$ which was known to be a deficiency of the original mild slope equation \cite{chamberlain1995modified}.  This shows that our  formulations for the modified mild slope equations are an accurate approximation for the MMSE as given in (\ref{eqnmmse}) or (\ref{eqnmmses}), but require far less computational effort due to their greater simplicity similar to the linear shallow water equations.

\subsection{Scattering over Artificial Bars }

\begin{figure}

 \vspace{15pt}
   
		\centering
	\includegraphics[width=0.7\textwidth]{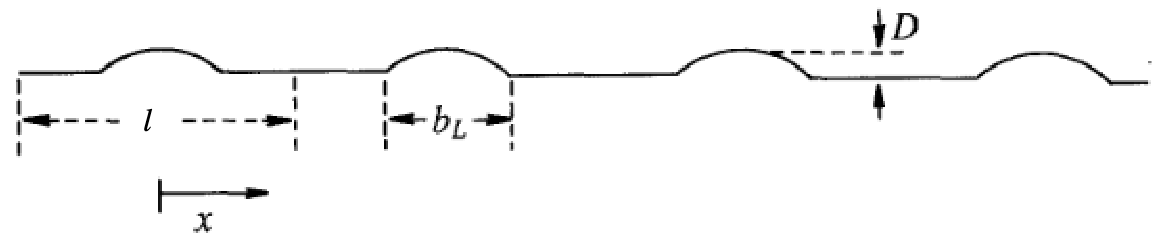}
 
	\caption{ Artificial bars in Kirby and Anton's scattering experiment. Sketch reproduced from \citet{kirby1990bragg} }	\label{fig:fig_kirbyanton}
\end{figure}

A variation of the scattering experiment by \citet{davies1984surface} has been carried out by \citet{kirby1990bragg} who placed  periodically spaced artificial bars of height ($\delta=\delta(x)>0$) on a flat surface of depth $h_0$.  In their study, the surface undulation  $\delta$ due to the artificial bars is given by a rectified cosine function as follows: 

\[
\delta(x) = 
\begin{cases}
D\cos\left(\frac{x}{b_{L}}(x-N\ell)\right) & \text{for } N\ell-\frac{b_{L}}{2} \leq x \leq N\ell+\frac{b_{L}}{2}\\
0 & \text{otherwise}
\end{cases}
\]
where $N=0,\;\ldots,\;N_{b}-1$. Due to the rectification of the bedform's cosine undulation, higher order harmonics beyond the original spatial  period $\ell$ of the artificial bars are being introduced, in contrast to the ripple bed topography of the experiment by \citet{davies1984surface} where the surface undulation only contains a single wave number. In the
experimental setup as sketched in Fig. \ref{fig:fig_kirbyanton}, the mean water depth was $h_0 = 15$cm. Height of the artificial bars   was $D=5$cm, which leads to a ratio
$D/h = 0.33$. This indicates a relatively large height of the bars  and could be the cause of some  deviations between predictions of mild slope equations  and experimental data. The bar length is $b_L = 50$cm. Two different spatial periods in the surface undulation were tested, namely $\ell=80\text{cm},120\text{cm}$,  corresponding
to cases where  higher harmonics in the bedform play a role of low importance and higher importance, respectively.\\

\begin{figure}

 \vspace{15pt}
   
		\centering
	\includegraphics[width=0.7\textwidth]{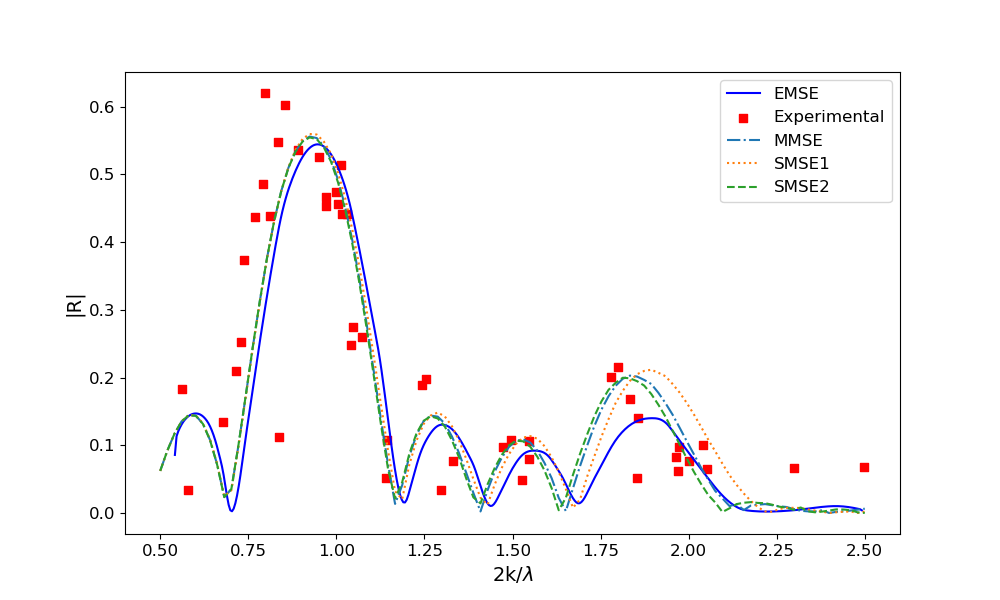}
 
	\caption{Reflection coefficient magnitude $|R|$ of scattering over rectified cosine bedform as a function of the incoming wave number $k$ normalized by the wave number $\lambda$ of the surface undulation. Experimental data by \citet{kirby1990bragg} for the case of 4 bars with spatial period $\ell=80\text{cm}$ compared to the MMSE and our simplified formulations (SMSE1), (SMSE2). 	} 	\label{fig:fig_kirbyanton1}
\end{figure}

Results for the case  $\ell=80\text{cm}$ are displayed in Fig. \ref{fig:fig_kirbyanton1}. The reflection coefficient magnitude $|R|$ is plotted against $2k/\lambda$, where $\lambda=2\pi/\ell$ is the wave number of the bedform. Experimental data from \citet{kirby1990bragg} are compared against predictions by the extended mild slope equation(EMSE)\cite{kirby1986general}, the MMSE, and our simplified formulations SMSE1 as well as SMSE2. It can be seen that all variants of the mild slope equations were able to capture the first dominant Bragg reflection at $2k\lambda=1$, whereas the MMSE and our simplified formulations were better at predicting the magnitude of the higher order resonance at $2k\lambda=2$ compared to EMSE. Again, we notice that despite the obvious simpler formulation of SMSE1 and SMSE2, they all showed excellent agreement with results obtained from MMSE.

\begin{figure}

 \vspace{15pt}
   
		\centering
	\includegraphics[width=0.7\textwidth]{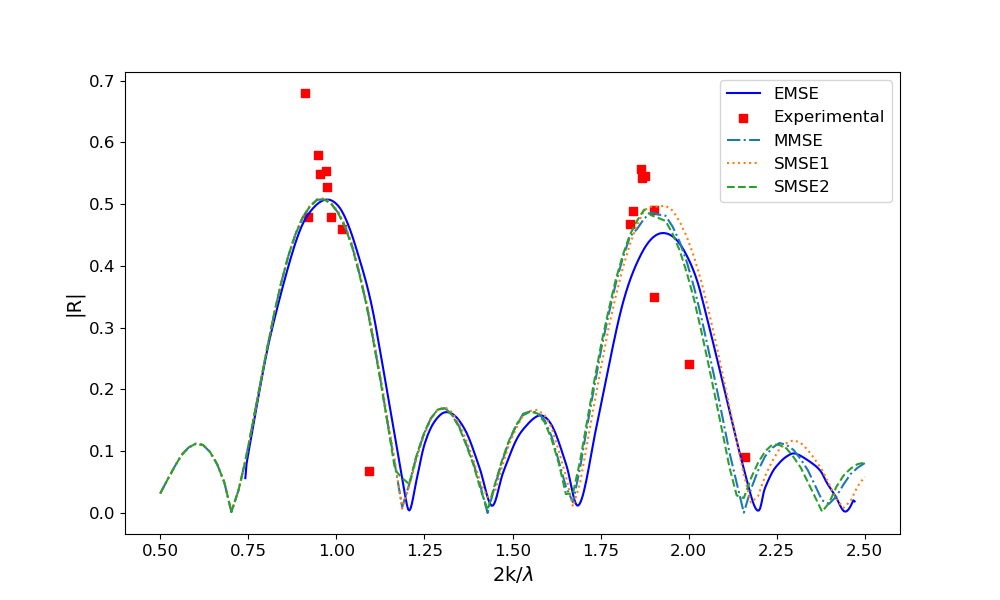}
 
	\caption{Reflection coefficient magnitude $|R|$ of scattering over rectified cosine bedform as a function of the incoming wave number $k$ normalized by the wave number $\lambda$ of the surface undulation. Experimental data by \citet{kirby1990bragg} for the case of 4 bars with spatial period $\ell=120\text{cm}$ compared to the MMSE and our simplified formulations (SMSE1), (SMSE2).} 	\label{fig:fig_kirbyanton2}
\end{figure}

Fig. \ref{fig:fig_kirbyanton2} displays reflection coefficient results for the second case where the spatial bar period is increased to  $\ell=120\text{cm}$. The same set of equations as before were used to compare against experimental data. Similar like the previous case, it is evident that all variants of the mild slope equations were again able to accurately predict the primary dominant Bragg reflection at $2k\lambda=1$. However,  the MMSE and our simplified formulations were again more accurate at capturing the magnitude of the higher order peak at $2k\lambda=2$ compared to EMSE which, despite its greater complexity, underestimated the reflection coefficient magnitude. For the range of incident wave numbers studied here,  the simpler formulations  SMSE1 and SMSE2 all showed nearly perfect agreement with results from MMSE, with  differences only starting to become noticeable at wave numbers $k>2.35$.

\section{Conclusions}
 
This study has successfully developed a simplified yet remarkably accurate approximation to the modified mild slope equations(MMSE) for modeling water wave scattering phenomena across a wide range of conditions. The newly derived mild slope equations (SMSE1,2) demonstrate that  mathematical simplicity does not need to come at the expense of physical accuracy as they achieve nearly identical results to the original MMSE for the most important scattering problems, while its reduced complexity offers significant practical advantages.

The main benefit of (SMSE1,2) lies in the fact that they do not require the exact evaluation of the nonlinear wave dispersion relation and many other related correction terms at each point, which is necessary in the oroginal MMSE. However, despite the clear resemblance of (SMSE1,2) to the linear shallow water equations (SWE),   it is capable of maintaining accuracy from shallow to intermediate water depths and across a broad spectrum of wavelengths. Hence, our newly developed equations  combine the best of two worlds: They share the same mathematical simplicity of the linear SWE while offering a wider range of validity beyond  the existing generalizations of SWE, which typically fail outside narrow parameter ranges. Through validation of multiple benchmark test cases for wave scattering problems,  we found that the simplified equations were able match the predictions by MMSE while eliminating unnecessary mathematical complexity.

The practical benefits of this simplification are manifold. First, the streamlined mathematical structure facilitates deeper theoretical analysis, making it possible to extract fundamental insights that were previously obscured by computational complexity. Second, the equations' elegant formulation enables the derivation of exact analytical solutions for problems that previously required numerical approximation - opening new avenues for understanding wave-topography interactions.
From an implementation perspective, the model's computational efficiency and compatibility with existing numerical schemes make it particularly valuable for both research and engineering applications..

Looking forward, this work establishes a foundation for several promising research directions. One important extension would be to generalize our  simplified mild slope equations to multiple dimensions, which appears to be a straightforward task. These novel equations can  also provide an ideal starting point for developing more efficient numerical methods for coastal wave prediction.

In conclusion, this study demonstrates that through the adoption of  simple approximations to the MMSE, we can achieve  both the accuracy and wide applicability of the MMSE  as well as the simplicity of the linear SWE. The resulting framework not only advances our theoretical understanding of wave scattering phenomena, but also provides engineers and researchers with a powerful new tool that combines physical fidelity with practical usability. Future work will explore applications of this approach to more complex wave environments and three-dimensional scattering problems.\\


\textbf{Declaration of Interests}: The authors report no conflict of interest.

\section*{Data Availability Statement}
The data that support the findings of this study are available
from the corresponding author upon reasonable request.


%
%

%


\bibliography{stability}

\end{document}